\documentclass[a4paper,11pt]{article}
\usepackage{pos}
\usepackage{subcaption}

\title{Electric Field Reconstruction with Information Field Theory}

\author*[a]{Simon Strähnz}
\author[a,b]{Tim Huege}
\author[c]{Philipp Frank}
\author[c]{Torsten Enßlin}

\affiliation[a]{Karlsruhe Institute of Technology,\\
  Kaiserstraße 12, 76131 Karlsruhe, Germany}

\affiliation[b]{Astrophysical Institute, Vrije Universiteit Brussel,\\
  Pleinlaan 2, 1050 Brussels, Belgium}

\affiliation[c]{Max-Planck-Institute for Astrophysics,\\
  Karl-Schwarzschild-Str. 1, 85748 Garching b. München, Germany}

\emailAdd{simon.straehnz@kit.edu}
\emailAdd{tim.huege@kit.edu}
\emailAdd{philipp@MPA-Garching.MPG.DE}
\emailAdd{ensslin@MPA-Garching.MPG.DE}

\abstract{
Reconstructing the electric field from the measured voltages in an antenna, unfolding the antenna response, comes with several problems. Due to the noisiness of the signal it is often necessary to disregard part of the bandwidth of the antenna. It is also not guaranteed, that this system of equations can be inverted at all. In any case, the noise of the measurement will be converted into noise on the electric field.

This could be solved by Bayesian inference, however, the electric field is continuous, which would lead to an infinite-dimensional latent space. Information field theory (IFT) has been developed to deal with this problem and allow for Bayesian reasoning on fields. It provides a theoretical backbone and effective tools to approach the inference as a discrete problem in the continuum limit, taking the continuous nature of fields into account.

We will present a first working signal model that can be used with IFT-based inference algorithms, which can successfully reconstruct the electric field. The model is based on the current understanding of air shower emission physics, modelling geomagnetic and charge-excess emission and their respective polarisation and spectra separately. Since Bayesian inference provides the posterior distribution, this method also gives an estimate on the uncertainty of the measured field. The performance of this method will be demonstrated with Monte-Carlo simulations of air shower radio signals.
}

\FullConference{10th International Workshop on Acoustic and Radio EeV Neutrino Detection Activities (ARENA2024)\\
11-14 June 2024\\
The Kavli Institute for Cosmological Physics, Chicago, IL, USA\\}

\newcommand{\Eamp}{\mathcal{E}}
\newcommand{\pol}{\text{pol}}
\newcommand{\dmax}{d_\text{max}}
\newcommand{\geo}{\text{geo}}
\newcommand{\ce}{\text{CE}}
\newcommand{\rfi}{\text{RFI}}

\begin{document}
\maketitle

\section{Introduction}
Measuring extensive air showers with radio detectors is a very attractive option. Radio detectors are cost-effective, so it is easy to cover a large area with them. They are also able to directly probe only the electromagnetic component of air showers, so they can be used in conjunction with particle detector arrays to measure the ratio of the electromagnetic to the muonic component, a crucial indicator to determine the type of the primary particle. The one big drawback to radio detection is its susceptibility to anthropogenic noise. The most power from air shower signals is expected in the range of 1 to 200 MHz, a frequency range not only very commonly used for radio communications, but also where the galactic background emission is very high.\ \cite{huegeRadio}

Reconstructing the extensive air shower from the measured voltages in radio antennas first requires the reconstruction of the electric field generated by the air shower at the location of the antenna. This is traditionally done by simply trying to invert the antenna response. In most detectors, antennas with only two polarisations are used, which comes with the limitation that it is impossible to infer a three-dimensional field from only a two-dimensional measurement. This problem can be circumvented by employing the far-field approximation (i.e., the electric field component in the direction of propagation of the wave vanishes). This still does not guarantee that a reconstruction is possible: If the direction of propagation is parallel to one polarisation of the antenna, it will not have any signal and the problem of dimensionality reappears. This method also does not take measurement noise into account. The noise on the measurement of the voltage in the antenna is simply propagated to noise on the measurement of the electric field.

A promising method to handle this reconstruction step is Bayesian inference. However, applying Bayesian inference to continuous fields is in general an ill-posed problem: A continuous field will have an infinite number of degrees of freedom, while the number of our measurements will always be finite. This can be solved with Information Field Theory.

\section{Information Field Theory}

Information Field Theory (IFT) studies, as the name suggests, information theory applied to fields. Through IFT it has been shown, that these ill-posed problems can be approached by the careful setting of prior probabilities using our current understanding of physics combined with Gaussian processes. In this way, it is possible to capture the physical and probabilistic constraints imposed on the field of interest and use them to handle infinite-dimensional latent spaces. Even though this requires discretisation at some point, it has been shown that this is allowable when working in the continuum limit. \cite{ensslinIFT}

In practice, working with IFT requires defining a suitable, probabilistic forward model, which includes the current understanding of the physical processes leading to our measurement, of the measurement process itself and of the prior probabilities associated with both. This model can then be used to apply Bayesian inference. Traditionally, there are two approaches to perform the inference: Markov-Chain Monte-Carlo sampling (MCMC) and Variational Inference (VI). The first approach (MCMC) directly samples the posterior distribution and is thus guaranteed to reconstruct the true posterior in the limit of infinite samples. However, it is also very computationally expensive and thus mostly unsuitable for IFT approaches, which work in the continuum limit and thus have a high number of degrees of freedom.

The more useful approach is to approximate the true posterior with a simpler but more easily calculable family of distributions, which is the basis of VI. Two VI methods have been developed for usage with IFT: Metric Gaussian VI (MGVI) \cite{knollmüller2020metricgaussianvariationalinference} uses a Gaussian distribution as the approximate posterior and estimates its covariance with the inverse Fisher metric. In an iterative process, the covariance matrix is estimated and the Kullback-Leibler divergence is minimised, until a satisfactory agreement is reached. In a further development of this method, Geometric VI (geoVI) \cite{frank2021geometricvariationalinference} can also handle non-Gaussian posteriors by way of applying a coordinate transform to the latent space to transform it into the space where the posterior is Gaussian.

\section{Forward Model}
The forward model used in this work for the variational inference was developed on the basis of the model used by C.~Welling~et~al.\ \cite{welling2021reconstructionnonrepeatingradiopulses}. In general, the same approach of modelling the electric field in the frequency domain as an amplitude ($\Eamp$) and phase ($\phi$) with a certain linear polarisation defined by a polarisation angle ($\phi_\pol$) is applied. The entire problem is considered under the far field approximation ($E_r = 0$):
\begin{equation}
    \vec{E}(f) = \Eamp(f) e^{i\phi(f)} (\cos\phi_\pol \hat{\theta} + \sin\phi_\pol \hat{\varphi})
\end{equation}
Here, geomagnetic emission and charge-excess emission have been modelled separately with their own separate amplitudes, phases, and polarisations. Narrowband RFI, as it is part of the measured electric field, has also been included in the model, however, this is modelled separately for every antenna polarisation.
\begin{align}
    \vec{E}(f) =& \omit\rlap{$\displaystyle \vec{E}_\geo(f) + \vec{E}_\ce(f) + \vec{E}_\rfi(f)$}\\
    =& [\Eamp_\geo(f) e^{i\phi_\geo(f)} \cos\phi_{\pol, \geo} +& \Eamp_\ce(f) e^{i\phi_\ce(f)} \cos\phi_{\pol, \ce} +& \Eamp_{\rfi, \theta}(f) e^{i\phi_{\rfi,\theta}(f)}] \hat{\theta}\\
    +&[\Eamp_\geo(f) e^{i\phi_\geo(f)} \sin\phi_{\pol, \geo} +& \Eamp_\ce(f) e^{i\phi_\ce(f)} \sin\phi_{\pol, \ce} +& \Eamp_{\rfi,\varphi}(f) e^{i\phi_{\rfi,\varphi}(f)}] \hat{\varphi}
\end{align}

\subsection{Detector model}
The measured data is assumed to be described by the equation
\begin{equation}\label{eq:measurement_equation}
    \vec{d}_i = \vec{V}(t_i) + \vec{n}_i
\end{equation}
i.e., the measured voltage in bin $i$ is the voltage $\vec{V}$ induced by the electric field in the antenna at the corresponding time $t_i$ plus some measurement noise $\vec{n}_i$.
The voltage is obtained by applying the instrument response to the electric field in frequency space and then transformed into real space with the inverse discrete Fourier transform:
\begin{equation}
    \vec{V}(t) = \text{DFT}^{-1}\left(\vec{V}(f)\right) = \text{DFT}^{-1}\left(\Tilde{H}_f \vec{E}(f)\right)
\end{equation}
Hereby is $\Tilde{H}_f$ the frequency dependent complex instrument response operator which is taken to be linear and given by
\begin{equation}\label{eq:matrix}
    \Tilde{H}_f = \left(\begin{matrix}
                    H_\theta^1 & H_\varphi^1 & H_r^1 \\
                    H_\theta^2 & H_\varphi^2 & H_r^2 \\
                    H_\theta^3 & H_\varphi^3 & H_r^3
                  \end{matrix}\right)_f
\end{equation}
where $H^i_x$ is the response of the $i$th channel to the electric field in the $\hat{x}$ direction. Since all $H^i_x$ are complex, this encodes the amplitude as well as the phase shift of the induced voltage.

The measurement equation \ref{eq:measurement_equation} can be rearranged as:
\begin{equation}
    n_i = \vec{d}_i - \vec{V}(t_i)
\end{equation}
Assuming that the noise in each bin $n_i$ is independently identically distributed according to some distribution $\mathcal{N}$, it follows that the likelihood of the measurement (i.e., the probability of measuring $\vec{d}_i$ given the signal $\vec{V_i}$ is given simply by:
\begin{equation}
    \mathcal{L} = P(\vec{d}_i | \vec{V}(t_i)) = \mathcal{N}(\vec{d}_i - \vec{V}(t_i))
\end{equation}

\subsection{Signal model}
Due to the limited bandwidth in many radio detectors, the amplitude ($\Eamp$) is modelled semi-parametrically: S.~Martinelli~et~al.\ \cite{martinelli2023parameterizationofthefrequencyspectrum} have shown that the amplitude of the spectrum of radio pulses from air showers can be described by linear~($L$) or quadratic~($Q$) exponentials
\begin{align}
    L &= A \cdot 10^{m_f(f-f_0)}\\
    Q &= A \cdot 10^{m_f(f-f_0) + m_{f2}(f-f_0)^2}
\end{align}
whereby the parameters $m_f$, $m_{f2}$ can be parameterised as a function the distance from the shower axis in units of the Cherenkov radius ($r/r_c$) and the geometric distance to shower maximum ($\dmax$). They have shown that the spectrum of the geomagnetic emission is best described using the quadratic exponential, whereas the charge-excess emission is already described well by the linear exponential. These parametrisations have been used as the parametric part of this forward model. To allow for deviations from this parametrisation, it is multiplied by an integrated Ornstein–Uhlenbeck process. This Gaussian process can be described as a biased, or bound, random walk. For this use case, it has been configured to create variations on the order of $10\%$ around 1.

\subsection{Noise model}
The noise is described separately as noise from narrowband RFI, which is then incorporated into the signal model and the remaining instrument and galactic noise, which is modelled as independently identically distributed variables according to the noise distribution $\mathcal{N}$. 

The narrowband RFI is modelled in frequency space and independently for each frequency bin. The amplitude is modelled as a random variable drawn from an inverse gamma distribution. Since this distribution falls off very quickly, this means that the amplitude is close to zero for almost all frequency bins, while there is still a reasonable possibility of very high peaks. 
The phase of the RFI is modelled as drawn from a wide Gaussian, making the prior practically agnostic on the phase of each RFI contribution

The remaining noise corresponding to the $n$ in equation \ref{eq:measurement_equation} is modelled as a multivariate Gaussian with a diagonal covariance matrix based on the standard deviation of the measured trace. This is derived for each measurement individually by calculating the sample standard deviation over a noise trace.
\begin{equation}
    \mathcal{N} = \mathcal{G}(0, \Sigma)
\end{equation}
\begin{equation}
    \Sigma^2 = \text{diag}(\sigma_s^2)
\end{equation}
This is by far the weakest link of the model and needs to be improved upon by better analysis of the noise distribution, which is currently ongoing: Since all detectors have a defined pass band, the noise is expected to be at least partly correlated. It has also already been shown that the real measurement noise is not purely thermal, i.e., it does not follow a normal distribution but has more pronounced tails. Thus, the current model severely underestimates the possibility of large deviations between signal and measurement due to noise.

\section{Performance on simulations}
This model has been tested on 800 CoREAS \cite{huegeCoREAS} simulations of inclined extensive air showers for proton, helium, nitrogen and iron primaries. These simulations range in energy from $10^{18.4}$~eV to $10^{20.1}$~eV, in zenith angle from $65^\circ$ to $85^\circ$ and over the entire azimuth angle range. The simulations were then passed through the detector simulation for the radio detector of the Pierre Auger Observatory \cite{huegeRd, radioOffline}. As a last step before reconstruction, measured noise from the Pierre Auger Observatory site was added to create realistic simulations of measurements. These were then reconstructed with the NIFTy framework \cite{niftyre} using the geoVI algorithm and the model as described above. For comparison, the full library (4000 showers) has been reconstructed using the current methods \cite{Schlueter2022_1000149113} and treated similarly. In the standard method, the antenna model is simply unfolded by inverting the antenna response matrix (eq. \ref{eq:matrix}) and applying it to the measured voltage. Further analysis in then usually performed on the energy fluence, the amount of energy deposited per unit area. This can be calculated directly from the measured E-Field when using the model described in this work. For the standard method, noise is handled in this step: The energy fluence is calculated by subtracting the power of the noise (estimated from a noise window) from the power from the pulse (calculated in a signal window).

\subsection{Reconstruction of the electric field}

\begin{figure}
    \centering
    \includegraphics[width=\linewidth]{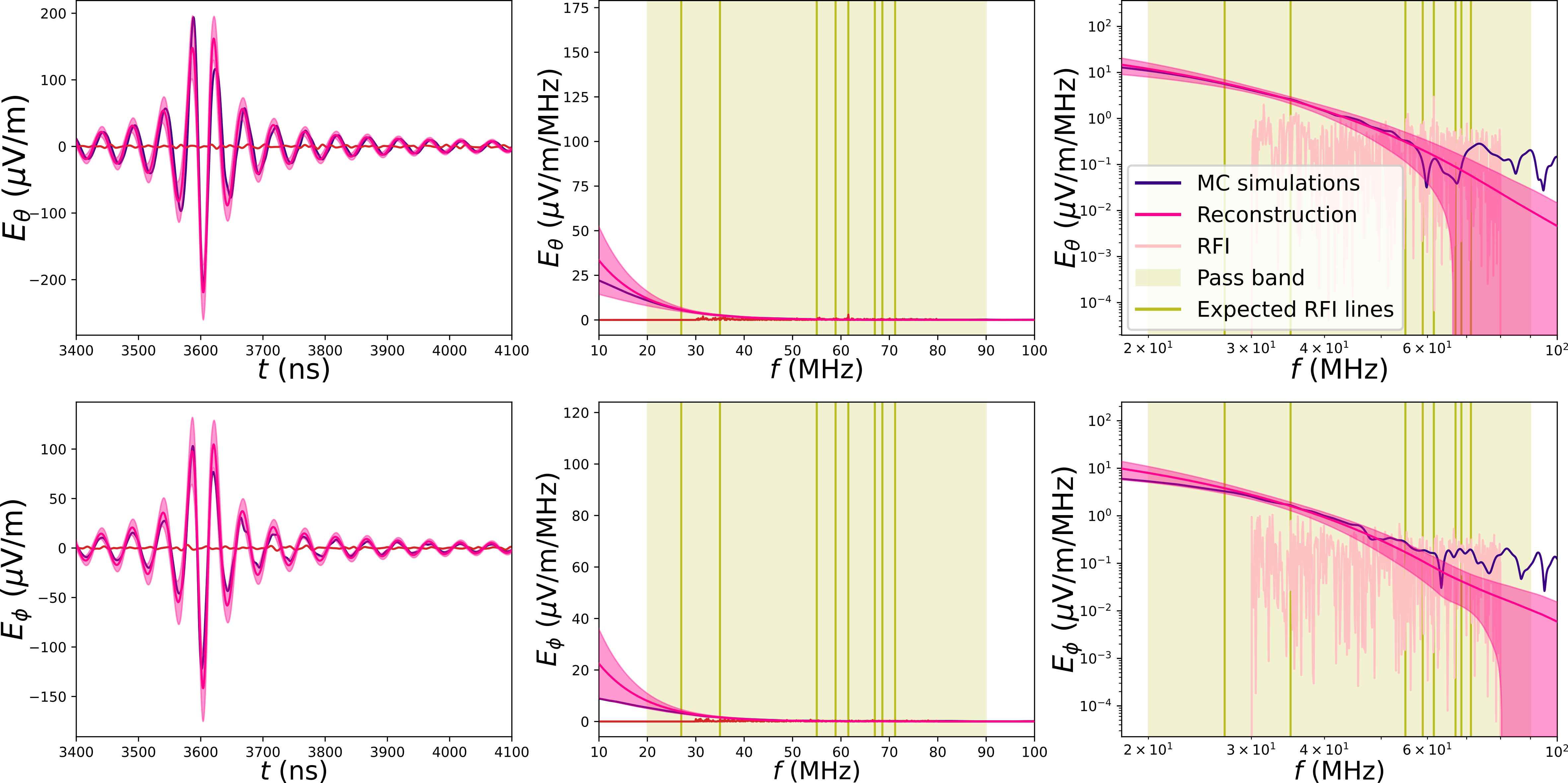}
    \caption{Example of the reconstruction of the electric field with the method presented in this work. Each row shows one polarisation. The first column shows the reconstruction in the time domain, zoomed in on the pulse. The middle and right columns show the amplitude spectrum in a linear and fully logarithmic plot, respectively. Expected RFI lines and reconstructed narrowband RFI are also shown (the reconstructed RFI is best seen in log-log coordinates, because of its intended small magnitudes)}
    \label{fig:exampleReco}
\end{figure}

An example of the improvement of the reconstruction of the electric field can be seen in figure~\ref{fig:exampleReco}. It shows the simulated cosmic ray pulse in the time domain and frequency domain and the reconstruction, including a 1$\sigma$ credible interval. This example is of a very-low signal radio detector station, which would usually have to be disregarded in further reconstruction. It can clearly be seen, that the IFT reconstruction is in very good agreement with the simulated ground truth. 

\begin{figure}
    \centering
    \begin{subfigure}{0.45\textwidth}
    \centering
        \includegraphics[width=\linewidth]{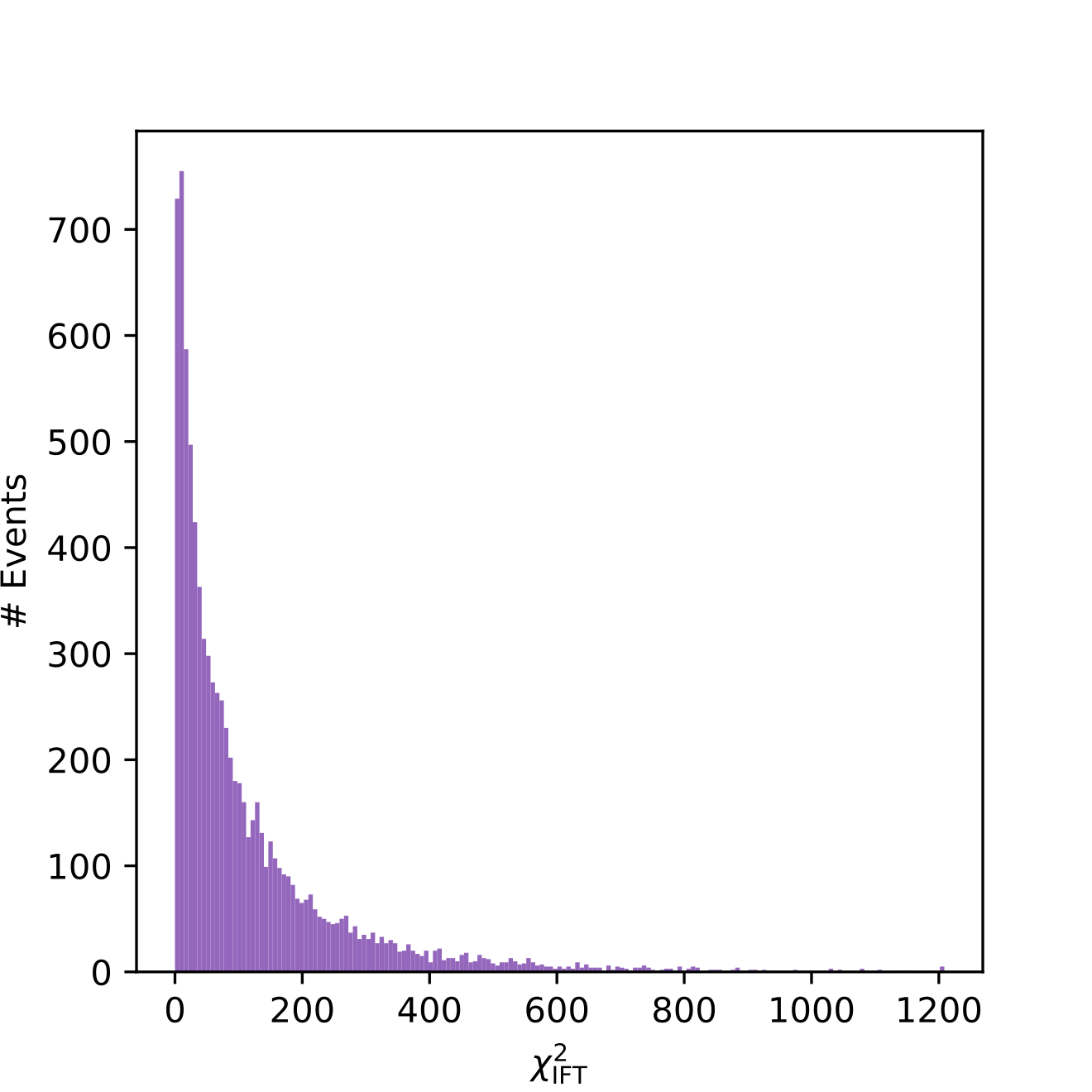}
        \caption{IFT}
        \label{fig:chi2:ift}
    \end{subfigure}
    \begin{subfigure}{0.45\textwidth}
        \centering
        \includegraphics[width=\linewidth]{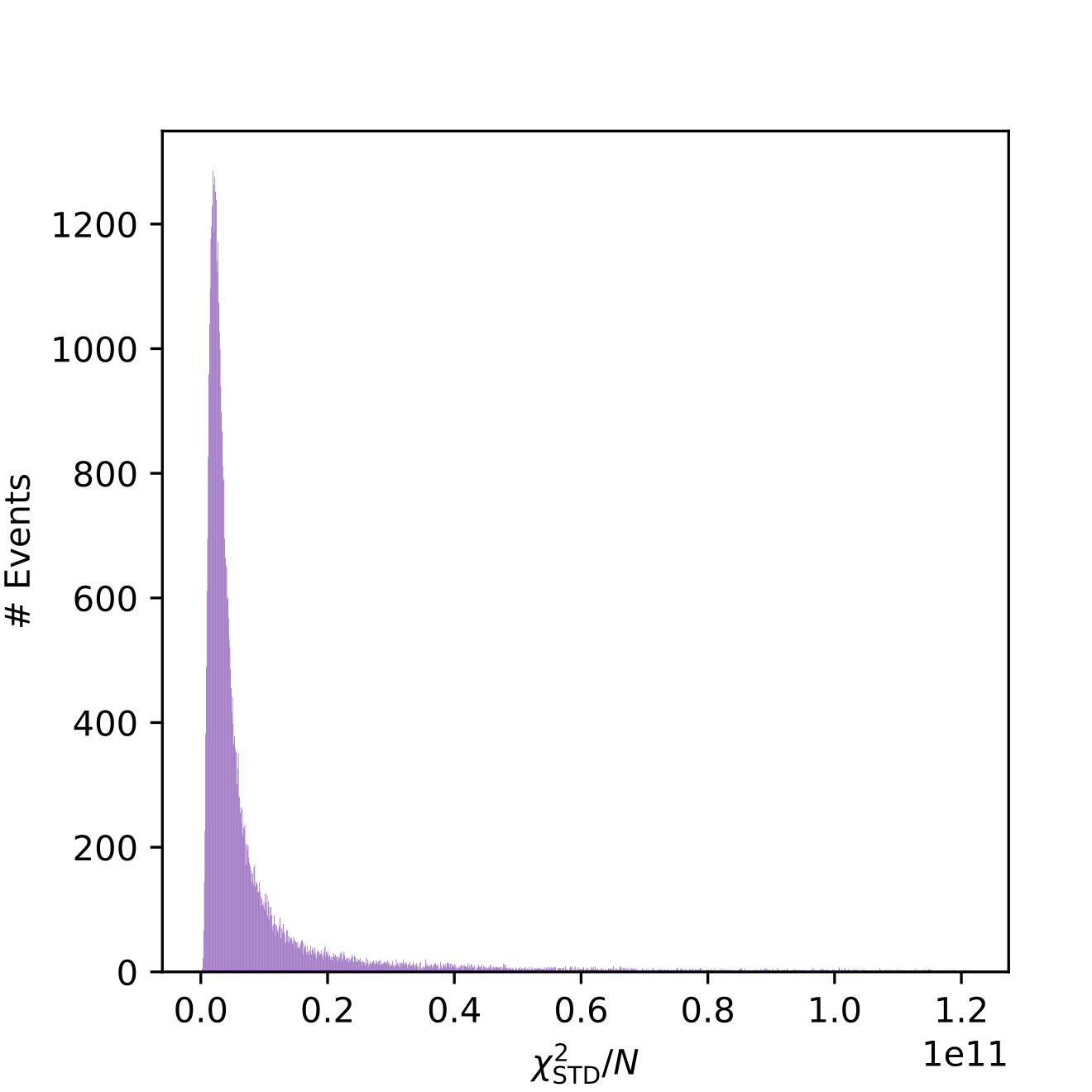}
        \caption{Standard}
        \label{fig:chi2:std}
    \end{subfigure}
    \caption{$\chi^2/N$ distribution for the comparison between simulations and reconstruction for the IFT method presented in this work (a) and the standard method of unfolding (b). Note that the axis for the standard method ranges to $1.2\cdot10^{11}$. The mean of the distribution for the IFT method is $163$, the mode is $6.5\pm0.9$. For the standard method, the mean is $1.8\cdot 10^{10}$ and the mode is $1.7\cdot10^9$.}
    \label{fig:chi2}
\end{figure}

To test the overall performance of the reconstruction of the electric field, a modified $\chi^2=\sum(x-\Bar{x})^2/\sigma_x^2$ has been calculated for all traces. For a normal least-squares fit, a $\chi^2$ score would be divided by the number of degrees of freedom, i.e., the number of data points minus the number of parameters. Since this method uses Bayesian inference and thus can have many more parameters than data points, this is not a reasonable approach.\footnote{In fact, this method employs roughly 4 times more parameters than data points, which are however strictly constrained by their prior distribution.} Instead, the quality of fit can be judged by dividing $\chi^2$ by the number of data points $N$. The distribution of this quantity should then be a $\chi^2$ distribution with a mean of 1 and a mode of $1-2/N$ if the errors are assumed to be the standard deviation of a normal distribution. The distributions for this reconstruction and the standard method can be seen in figure \ref{fig:chi2}. For the standard method, the naive assumption has been made, that the sample standard deviation of the noise can be used as an approximate error on the electric field reconstruction. From the distributions, it is immediately obvious, that the IFT-based reconstruction outperforms the standard reconstruction by a lot. This is easily explained by the fact that the new method can deal with measurement noise, while the old method simply propagates this noise to the electric field. However, the statistics of the new method (the mean of the distribution is $\approx163$, the mode is $\approx6.5$) indicate that the errors are still not well understood and underestimated. The algorithm used in reconstruction (geoVI \cite{frank2021geometricvariationalinference}), although being able to handle non-Gaussian posterior distributions, is still only an approximation and can underestimate the variance if the posterior is unfavourably shaped. Especially in cases like this, where the noise model is simplistic and underestimates the tails of the noise distribution. That said, it is still a massive improvement compared to the old method.

\subsection{Reconstruction of the energy fluence}
\begin{figure}
    \centering
    \begin{subfigure}{0.49\textwidth}
        \includegraphics[width=\linewidth]{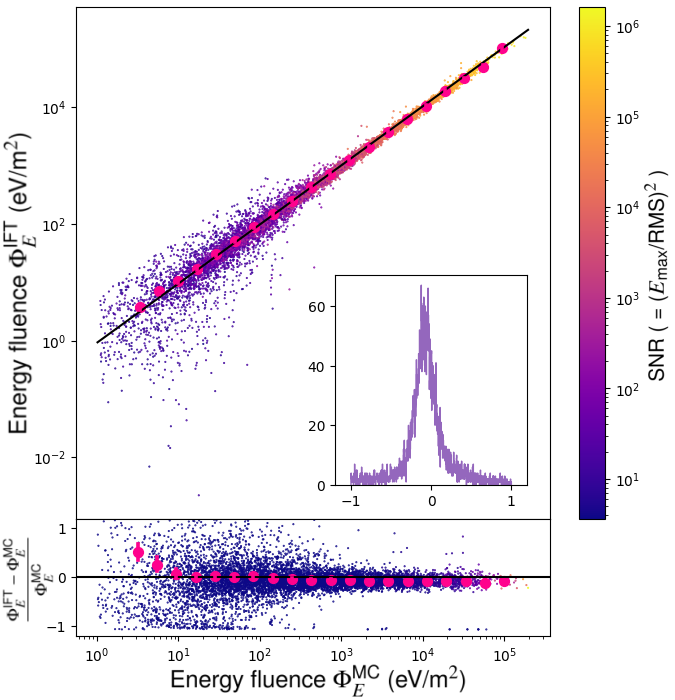}
        \caption{IFT}
        \label{fig:fluence:ift}
    \end{subfigure}
    \begin{subfigure}{0.49\textwidth}
        \includegraphics[width=\linewidth]{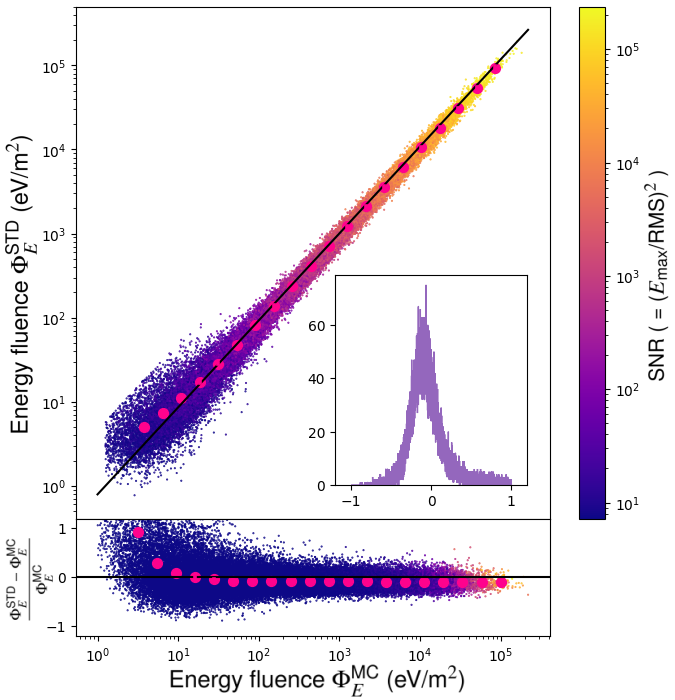}
        \caption{Standard}
        \label{fig:fluence:std}
    \end{subfigure}
    \caption{Comparison of the reconstructed to the simulated energy fluence for the IFT and the standard reconstruction method. The inset shows the distribution of the residuals. The profile dots and errors bars (mostly invisible) show the mean and standard deviation of the datapoints.}
    \label{fig:fluence}
\end{figure}

A comparison of the reconstructed energy fluence (the amount of energy deposited per unit area, proportional to the time-integral of the squared electric field) can be seen in figure \ref{fig:fluence}. Both methods of handling the noise seem to be working equally well, at least for large signals. For smaller signals, the IFT method introduces slightly less bias than the standard method. This is easily explained by the reduced influence of noise in the IFT method. While the total measured power becomes dominated by noise at some point for the standard method, the method presented here does either fail to perform inference, or find the signal pulse without noise. Still, there is no large benefit of performing the computationally heavy IFT reconstruction if the further analysis is based on the energy fluence alone. The radio detector stations for which the calculation is less biased have signals so small that they would not be used in a reconstruction, and even if they were used, their contribution to the reconstruction would be negligible.

\section{Conclusion}
A forward model for the radio emission of extensive air showers, based on our current understanding of emission processes, has been developed and successfully implemented in an Information Field Theory based reconstruction algorithm to infer the electric field from noisy measurements. The reconstruction method has been tested on Monte-Carlo simulations and compared to the standard unfolding method. Based on the $\chi^2/N$ distributions, it has been shown, that the reconstruction of the electric field is greatly improved. When comparing the energy fluence, a quantity which is the basis of most further analysis methods currently in use, the IFT-based method does not offer a great improvement except for antennas with very small signals. This means, that current analysis methods can not benefit from the improved e-field reconstruction. However, there are further reconstruction methods under development, which depend on the pulse shape rather than the power alone (see~e.g.~\cite{Karastathis:20230H}), which are relatively sensitive to noise and could be used to better effect with the electric field reconstruction detailed in this work. In the future, this method shall be developed further by using a single event-level model of the electric field, reconstructing the signal from all antennas at once.

\acknowledgments
Simon Strähnz and Philipp Frank acknowledge funding through the German Federal Ministry of Education and Research for the project ErUM-IFT: Informationsfeldtheorie für Experimente an Großforschungsanlagen (Förderkennzeichen: 05D23EO1).

\bibliographystyle{JHEP}
\footnotesize
\bibliography{bibliography}


\end{document}